\documentclass[aps,prd,preprint
,tightenlines,letterpaper,
nofootinbib,showpacs]{revtex4-1}
\usepackage{amssymb,latexsym}
\usepackage{amsmath,amsbsy,bbm}
\usepackage{epsfig,bm}
\usepackage{graphicx,comment}
\DeclareMathOperator{\tr}{tr}

\begin{document}

\def\a{{\alpha}}
\def\b{{\beta}}
\def\d{{\delta}}
\def\D{{\Delta}}
\def\X{{\Xi}}
\def\e{{\varepsilon}}
\def\g{{\gamma}}
\def\G{{\Gamma}}
\def\k{{\kappa}}
\def\l{{\lambda}}
\def\L{{\Lambda}}
\def\m{{\mu}}
\def\n{{\nu}}
\def\o{{\omega}}
\def\O{{\Omega}}
\def\S{{\Sigma}}
\def\s{{\sigma}}
\def\th{{\theta}}

\def\ol#1{{\overline{#1}}}

\def\Dslash{D\hskip-0.65em /}
\def\Dtslash{\tilde{D} \hskip-0.65em /}

\def\CPT{{$\chi$PT}}
\def\QCPT{{Q$\chi$PT}}
\def\PQCPT{{PQ$\chi$PT}}
\def\tr{\text{tr}}
\def\str{\text{str}}
\def\diag{\text{diag}}
\def\order{{\mathcal O}}

\def\cF{{\mathcal F}}
\def\cS{{\mathcal S}}
\def\cC{{\mathcal C}}
\def\cB{{\mathcal B}}
\def\cT{{\mathcal T}}
\def\cQ{{\mathcal Q}}
\def\cL{{\mathcal L}}
\def\cO{{\mathcal O}}
\def\cA{{\mathcal A}}
\def\cQ{{\mathcal Q}}
\def\cR{{\mathcal R}}
\def\cH{{\mathcal H}}
\def\cW{{\mathcal W}}
\def\cM{{\mathcal M}}
\def\cD{{\mathcal D}}
\def\cN{{\mathcal N}}
\def\cP{{\mathcal P}}
\def\cK{{\mathcal K}}
\def\Qt{{\tilde{Q}}}
\def\Dt{{\tilde{D}}}
\def\St{{\tilde{\Sigma}}}
\def\cBt{{\tilde{\mathcal{B}}}}
\def\cDt{{\tilde{\mathcal{D}}}}
\def\cTt{{\tilde{\mathcal{T}}}}
\def\cMt{{\tilde{\mathcal{M}}}}
\def\At{{\tilde{A}}}
\def\cNt{{\tilde{\mathcal{N}}}}
\def\cOt{{\tilde{\mathcal{O}}}}
\def\cPt{{\tilde{\mathcal{P}}}}
\def\cI{{\mathcal{I}}}
\def\cJ{{\mathcal{J}}}

\def\eqref#1{{(\ref{#1})}}


\preprint{UMD-40762-479}
\preprint{INT-PUB-10-023}

 \author{B.~C.~Tiburzi}
\email[]{$\texttt{bctiburz@umd.edu}$}
\affiliation{%
Maryland Center for Fundamental Physics, 
Department of Physics, 
University of Maryland, 
College Park,  
MD 20742-4111, 
USA
}

\title{Chiral Lattice Fermions, Minimal Doubling, \\
and the Axial Anomaly} 

\begin{abstract}
Exact chiral symmetry at finite lattice spacing would preclude the axial anomaly. 
In order to describe a continuum quantum field theory of Dirac fermions, 
lattice actions with purported exact chiral symmetry must break the flavor-singlet axial symmetry. 
We demonstrate that this is indeed the case by using a minimally doubled fermion action. 
For simplicity we consider the Abelian axial anomaly in two dimensions. 
At finite lattice spacing and with gauge interactions, 
the axial anomaly arises from non-conservation of the flavor-singlet current.
Similar non-conservation also leads to the axial anomaly in the case of the na\"ive lattice action. 
For minimally doubled actions, 
however,
fine tuning of the action and axial current is necessary to arrive at the anomaly. 
Conservation of the flavor non-singlet vector current additionally requires the current to be fine tuned. 
Finally we determine that the chiral projection of a minimally doubled fermion action 
can be used to arrive at a lattice theory with an undoubled Dirac fermion possessing the correct anomaly in the continuum limit.
\end{abstract}


\maketitle

\section{Introduction}                                                                           %

The study of strongly interacting gauge theories has made remarkable progress due to 
numerical computations on Euclidean space-time lattices.  
Lattice gauge theory provides a gauge invariant regularization of vector-like theories, 
such as QCD. 
Inclusion of a desired number of light fermionic degrees of freedom, 
however,
is not a simple task. 
The na\"ive discretization, 
$D$, 
of the massless free fermion Dirac operator maintains exact chiral symmetry,
$\{ D^{-1}, \gamma_5 \} = 0$, 
but produces fermion doubling.
For each na\"ive fermion in 
$d$-dimensions, 
$2^d$ 
Dirac fermions emerge in the continuum limit. 
Under some reasonable assumptions, 
one can make this result more general.
The Nielsen-Ninomiya theorem%
~\cite{Nielsen:1981hk,*Nielsen:1980rz,*Nielsen:1981xu}
essentially states that chiral symmetry must be sacrificed to avoid doubling. 
This was the approach taken in the pioneering work of Wilson%
~\cite{Wilson:1974sk}.
The Wilson Dirac operator, 
$D_W$,  
avoids fermion doubling because it contains an irrelevant operator that breaks chiral symmetry, 
so that one has~$\{ D_W^{-1}, \gamma_5 \} \neq 0$. 
At some level, 
chiral symmetry must be broken by all lattice actions with any hope of describing Dirac fermions in the continuum limit.
The reason is that exact chiral symmetry precludes the flavor-singlet axial anomaly. 
For the Wilson action, 
chiral symmetry breaking introduced by the Wilson term leads to the correct axial anomaly in the continuum limit%
~\cite{Karsten:1980wd}
(we provide a simple derivation in the Appendix).

Rather elegant solutions to the fermion doubling problem exist. 
Starting from continuum QCD and smearing the fermion fields around lattice sites, 
one can deduce the properties of an ideal lattice Dirac operator, 
$D_{GW}$.
Such an operator satisfies what is known as the Ginsparg-Wilson relation%
~\cite{Ginsparg:1982bj}, 
namely
$\{ D^{-1}_{GW}, \gamma_5 \} = \gamma_5$.
While Ginsparg-Wilson fermions do not have the usual continuum chiral symmetry, 
they maintain a lattice version of chiral symmetry, 
which guarantees good chiral symmetry properties of the theory; 
but, at the same time, ensures that the correct axial anomaly is recovered%
~\cite{Luscher:1998pqa}.
Domain wall fermions%
~\cite{Kaplan:1992bt},
and their four-dimensional reduction, 
overlap fermions%
~\cite{Narayanan:1993ss},
are explicit solutions to the Ginsparg-Wilson relation.%
\footnote{
The Ginsparg-Wilson relation can also be realized through the construction of perfect lattice actions%
~\cite{Hasenfratz:1998ri}. 
} 
While elegant, 
these fermions are numerically costly. 
It is of great interest to have inexpensive chiral fermions for QCD computations,
and additionally for studies of strongly coupled dynamics in theories beyond the standard model, 
for a review of such theories, see%
~\cite{Hill:2002ap}.

The 
na\"ive action has a form of chiral symmetry not encompassed by the Ginsparg-Wilson relation. 
As the low-energy modes emerge from several regions within the Brillouin zone, 
the flavors of na\"ive fermions 
(often referred to as tastes) 
are non-local in space-time.%
\footnote{
While there is no unique way to define the flavors of na\"ive fermions, 
all choices are non-local and differ by lattice-spacing artifacts.  
The mildest non-locality possible has a range of just one lattice spacing. 
For this choice, 
the relative coordinates within an elementary hypercube can be regarded as internal degrees of freedom which give rise to the flavors.
Rewritten in this way, 
the action becomes local and contains flavor-breaking interactions. 
} 
While the free na\"ive action maintains an exact flavor-singlet axial symmetry, 
the corresponding axial transformation of the fermion field is non-local.
In the presence of gauge interactions,  
the flavor-singlet axial symmetry must be broken. 
Consequently the correct axial anomaly is recovered in the continuum, 
as was realized in the early days of lattice QCD%
~\cite{Sharatchandra:1981si}. 
Today one commonly encounters the statement that na\"ive fermions do not have an axial anomaly, 
see, 
for example,%
~\cite{Gupta:1997nd,*Chandrasekharan:2004cn,*Kaplan:2009yg}.
Such statements are solely based on the local axial transformation, 
$\psi_x \to \exp ( i \theta \gamma_5 ) \psi_x$, 
which is a singlet transformation of the na\"ive field.
This transformation generates an exact symmetry of the interacting theory;  
but, 
in terms of a flavor interpretation, 
careful consideration reveals that it is non-singlet%
~\cite{Karsten:1980wd,Sharatchandra:1981si}.

Here we detail the way in which the axial anomaly emerges for 
theories with chiral symmetry of the form $\{ D^{-1}, \gamma_5 \} = 0$. 
In our study, 
we use the minimally doubled fermion action of Wilczek~\cite{Wilczek:1987kw}.%
\footnote{ 
A very similar action appears to be first mentioned in~\cite{Karsten:1981gd}, 
where Nielsen is credited for suggesting it.
}
This simplifies the flavor content of the continuum theory
compared to the na\"ive fermion action, 
and consequently 
simplifies the discussion of flavor singlet and non-singlet currents. 
Recently interest has been renewed in minimally doubled 
fermion actions due to graphene-inspired constructions%
~\cite{Creutz:2007af,*Borici:2007kz,*Bedaque:2008jm}.
These actions, 
however, 
maintain fewer symmetries over the Wilczek action%
~\cite{Pernici:1994yj,*Bedaque:2008xs,*Capitani:2009yn,*Capitani:2010nn,*Creutz:2010cz},
giving us another reason to focus on it.
As with the na\"ive action, 
the local axial transformation, 
$\psi_x \to \exp ( i \theta \gamma_5) \psi_x$, 
generates a conserved flavor non-singlet current. 
On the other hand, 
the flavor-singlet axial transformation of the fermion field is inherently non-local
(symmetry transformations for minimally doubled fermions are collected in Table~\ref{t:Sym}).  
In the presence of gauge interactions, 
the corresponding singlet current is not conserved,
and is the source of the chiral anomaly.
An important additional finding of our study is that the non-singlet vector current is not conserved.
Current operator must be fine tuned in order to recover
conservation of the non-singlet vector current in the continuum limit. 
A similar malady affects the divergence of the singlet axial current;
hence, 
fine tuning is also required to arrive at the chiral anomaly. 
We also analyze the chiral projection of the Wilczek action, 
and demonstrate how a single Dirac fermion emerges from its 
continuum limit.

We organize our presentation as follows. 
The free minimally doubled action is spelled out in Sec.~\ref{s:MFD}, 
along with its symmetries, 
both discrete and continuous,
and the form of the conserved 
vector and axial-vector currents. 
The inclusion of gauge interactions is investigated in Sec.~\ref{s:VAVC}. 
When gauge interactions are included, 
the flavor-singlet axial-vector current is no longer conserved.
We use this observation to calculate the chiral anomaly in Sec.~\ref{s:CA}, 
where,  
for simplicity, 
we consider only the two-dimensional case.
The flavor non-singlet vector symmetry is also violated by gauge interactions.
We show that conservation of the corresponding current requires fine tuning of a marginal operator.
There is a similar such marginal operator for the flavor-singlet axial current, 
and it must be tuned to arrive at the correct chiral anomaly.
Chiral projection is taken up in Sec.~\ref{s:MDWF}, 
where we consider the minimally doubled action for two-component Weyl fermions. 
In the continuum limit, 
we show how this theory can describe an undoubled Dirac fermion with the correct chiral anomaly. 
The Appendix gives a simple derivation of the chiral anomaly for the case of the Wilson lattice action.
Finally, 
we summarize our findings in Sec.~\ref{s:S}.

\section{Minimally Doubled Fermions}                           
\label{s:MFD}                %

\subsection{Free action and its symmetries} %

To begin our discussion,
we consider a free lattice theory of minimally doubled fermions. 
In the absence of gauge interactions, 
the Wilczek action is given by
\begin{eqnarray} 
S 
&=& 
\frac{1}{2}
\sum_{x, \mu}
\Big[
\ol \psi_x
\gamma_\mu 
\psi_{x + \mu} 
-
\ol \psi_{x + \mu}
\gamma_\mu 
\psi_x 
\Big]
- 
\frac{i \lambda}{2} 
\sum_{x, j}
\Big[
\ol \psi_x
\gamma_4 
\psi_{x + j} 
+ 
\ol \psi_{x + j}
\gamma_4 
\psi_x 
-
2
\ol \psi_x 
\gamma_4
\psi_x
\Big]
\label{eq:action}
,\end{eqnarray}
where sums over 
$\mu$ 
run over all space-time axes, 
sums over 
$j$ 
run only over spatial axes, 
and sums over 
$x$ 
run over all lattice sites. 
Above, 
$\lambda$
is a real-valued parameter irrelevant to the low-energy dynamics
provided that the condition
$| \lambda | > \frac{1}{2}$
is met.

In momentum space, 
we define
$\psi(k) = \sum_{x} e^{ - i k \cdot x} \psi_x$, 
which renders the action in diagonal form,
with the free Dirac operator given by
\begin{eqnarray}
D^{(0)}(k) 
&=& 
\sum_\mu i \gamma_\mu \sin k_\mu
- 
i \gamma_4
\lambda
\sum_j   ( \cos k_j - 1)
\label{eq:free}
.\end{eqnarray}
There are precisely two zeros of this action for 
$| \lambda | > \frac{1}{2}$, 
namely at the momenta
$k^{(1)}_\mu = (0,0,0,0)$,
and 
$k^{(2)}_\mu = (0,0,0,\pi)$. 
Fields with restricted energy%
\footnote{
Throughout,
we shall generically use energy to refer to the fourth component of momentum. 
} 
correspond to quark flavors. 
We choose the partition based on the ball,%
\footnote{
The partition of the Brillouin zone into flavors is not unique, 
and alternate choices differ by lattice artifacts. 
Our partition is chosen in order to treat the flavors symmetrically. 
} 
$\mathcal{B} = \{ k_\mu:  | k_4 | < \frac{\pi}{2} \}$,  
so that
\begin{equation} \label{eq:1}
\psi(k) \Big|_{k_\mu \in \mathcal{B}} = \psi^{(1)} (k)
.\end{equation} 
On the other hand, 
careful consideration yields the second quark flavor%
\footnote{
In the free theory, 
one is not required to define the second flavor in exactly this way. 
Such freedom, however, is lifted by the inclusion of gauge interactions, 
and 
Eq.~\eqref{eq:2} is the only possibility allowing for both fermion flavors to encounter the same gauge field. 
} 
\begin{equation} \label{eq:2}
\psi(k) \Big|_{k_\mu \notin \mathcal{B}} = \gamma_4 \gamma_5 \psi^{(2)} (T_{\pi4} k)
,\end{equation}
with shifted momentum given by 
$T_{\pi4} k_\mu = \left(\bm{k},  \, k_4 + \pi \mod 2 \pi \right)$.

We can write the momentum-space action in terms of quark flavors
using the partition in Eqs.~\eqref{eq:1} and \eqref{eq:2}. 
Defining the isospinor 
$\Psi(k) = \begin{pmatrix} \psi^{(1)} (k) \\ \psi^{(2)} (k) \end{pmatrix}$,
the flavored action takes the form
\begin{eqnarray}
S 
&=&
\int_{\mathcal{B}} 
\frac{d^4k}{( 2\pi)^4} 
\ol \Psi (k)
\left[
\sum_\mu i  (\gamma_\mu \otimes 1) \sin k_\mu 
\right.
- i \lambda
\left.
(\gamma_4 \otimes \tau^3)
\sum_j 
( \cos k_j - 1) 
\right]
\Psi (k)
.\end{eqnarray}
Notice the momentum is restricted to 
$\mathcal{B}$, 
and we employ a 
$( \text{spin } \otimes \text{ flavor} )$ 
notation reminiscent of staggered fermions. 
In addition to cubic invariance, 
the free action has the symmetries listed in Table~\ref{t:Sym}.
%
%
%
%
\begin{table}[t]
\caption{%
Symmetries of the free minimally doubled action. 
Listed are the discrete and continuous symmetries of the action, 
and the corresponding transformations of the momentum-space and coordinate-space fermion fields.
The spinor representations for the discrete symmetries are:
$C = \gamma_2 \gamma_4$, 
$P = \gamma_4$, 
and
$T = \gamma_4 \gamma_5$, 
while the three-vector 
$\bm{1} = ( 1, 1, 1)$.
Site reflection is an antilinear transformation, 
whereas the remaining transformations are all linear. 
}
\label{t:Sym}
\begin{center}
\begin{tabular}{l||c|c}
Free Theory Symmetry &
Momentum Space &
Coordinate Space
\\
&
$\Psi(k) \longrightarrow$ &
$\psi_x \longrightarrow $
\tabularnewline
\hline
\hline
Site Reflection 
&
$\quad  e^{ - i \sum_j k_j} (T \otimes 1) \ol \Psi {}^T(\bm{k}, - k_4) \quad$ &
$\quad T \, \ol \psi {}^T_{\bm{1} - \bm{x},  x_4 } \quad$ 
\\
$P$
&
$(P \otimes 1) \Psi(-\bm{k}, k_4)$ &
$\quad P \left( \delta_\mathcal{B} \, \psi_{- \bm{x}, x_4} - \delta_{\overline{\mathcal{B}}} \, \psi_{- \bm{x}, x_4} \right) \quad$
\\
$CT$
&
$(C T \otimes 1) \ol \Psi {}^T (\bm{k}, - k_4)$ &
$C T \left( \delta_\cB \, \ol \psi {}^T_{\bm{x}, - x_4} - \delta_{\ol \cB} \, \ol \psi {}^T_{\bm{x}, - x_4} \right) $
\\
\hline
$T \, \times$ Flavor Rotation 
&
$\quad ( T \otimes i \cos \theta \, \tau^1 + i \sin \theta \, \tau^2) \Psi( \bm{k}, - k_4) \quad$ &
$i e^{ i \pi x_4} \left(  e^{ - i \theta} \delta_\cB \, \psi_{\bm{x}, -x_4} -  e^{ i \theta} \delta_{\ol \cB} \, \psi_{\bm{x}, -x_4} \right)$
\\
$U(1)$ Isosinglet Vector 
&
$e^{ i \theta ( 1 \otimes 1)} \Psi(k)$
&
$ e^{ i \theta} \psi_x$
\\
$U(1)$
Isovector Axial
&
$e^{ i \theta ( \gamma_5 \otimes \tau^3)} \Psi(k)$
&
$e^{i \theta \gamma_5} \psi_x$ 
\\
$U(1)$
Isovector Vector
&
$e^{i \theta (1 \otimes \tau^3)} \Psi(k)$
&
$e^{ i \theta} \delta_{\mathcal{B}} \, \psi_x 
+ 
e^{ - i \theta} \delta_{\overline{\mathcal{B}}} \, \psi_x $
\\
$U(1)$ Isosinglet Axial
&
$e^{i \theta (\gamma_5 \otimes 1)} \Psi(k)$
&
$e^{ i \theta \gamma_5} \delta_{\mathcal{B}} \, \psi_x 
+ 
e^{ - i \theta \gamma_5} \delta_{\overline{\mathcal{B}}} \, \psi_x$
\end{tabular}
\end{center}
\end{table}
The continuous vector and axial symmetries
can alternately be described as a chiral symmetry:
$U(1)_L \otimes U(1)_R$. 
There is such chiral symmetry for the isosinglet, 
$1$, 
and isovector,
$\tau^3$, 
flavor combinations.

In the table, 
we have also spelled out how the coordinate space field, 
$\psi_x$, 
transforms under the momentum-space symmetries.
The 
$P$, 
$CT$, 
$T \, \times$ 
flavor rotation,
isovector vector, 
and isosinglet axial transformations are non-local in time.
They depend on the time-smeared fields%
\footnote{
The temporal non-locality present in 
$\delta_\cB \, \psi_x$
is of maximal range.
A milder form of non-locality is needed in practice, 
and would result from a different definition of the flavors.  
For example, 
the infinitesimal transformation,
$\psi_x \longrightarrow \psi_x +  \frac{i}{2}  \theta \gamma_5 ( \psi_{x + \hat{4}} + \psi_{x- \hat{4}} ) + \ldots \, $,
corresponds to a isosinglet axial transformation different than that in Table~\ref{t:Sym}.
While this transformation does not exponentiate simply, it is minimally non-local in time.  
}
\begin{equation}
\delta_{\mathcal{B}} \, \psi_x
= 
\int_{- \frac{\pi}{2}}^{+ \frac{\pi}{ 2}} 
\frac{d k_4}{2 \pi}
\sum_{y_4}
e^{ i k_4 (x_4 - y_4) }
\psi_{\bm{x}, y_4}
\label{eq:smear}
,\end{equation} 
and similarly for 
$\delta_{\overline{\mathcal{B}}} \, \psi_x$, 
for which the energy integration is over the complement, 
namely
$\overline{\mathcal{B}}  = \{ k_4:  |k_4| \nless \frac{\pi}{2} \}$. 
It is straightforward to verify that the coordinate space transformations in 
Table~\ref{t:Sym} 
leave the action in 
Eq.~\eqref{eq:action}
invariant. 
Potential symmetry violating terms from the non-local transformations
identically vanish because the momentum regions are non-overlapping,
$\mathcal{B} \cap \overline{\mathcal{B}} = \{ \}$.

\subsection{Vector and Axial-Vector Currents}                                    %

For a generic current, 
$\mathcal{J}_\mu(x)$,
the lattice conservation law is precisely the statement:
$\sum_\mu \nabla^*_\mu \, \mathcal{J}_\mu(x) = 0$, 
where 
$\nabla^*_\mu$ 
is the backwards difference operator, 
$\nabla^*_\mu \, f_x \equiv f_x - f_{x- \mu}$. 
To derive the conserved currents, 
we appeal to the Noether procedure, 
which yields four conserved point-split currents.
The two local symmetry transformations yield
the isosinglet vector current
\begin{eqnarray}
J_\mu(x) 
&=& 
\frac{1}{2}
\Big[
\ol \psi_x \gamma_\mu \psi_{x + \mu}
+
\ol \psi_{x + \mu} \gamma_\mu \psi_x
- 
i \lambda   \, \delta_{\mu j}
\Big(
\ol \psi_x \gamma_4 \psi_{x + j}
-
\ol \psi_{x + j} \gamma_4 \psi_x
\Big)
\Big], \label{eq:isosingletvector}
\end{eqnarray}
and
the isovector axial current,
$J^3_{\mu 5}(x)$, 
which is given by the same expression as
$J_\mu(x)$, 
but with the replacement:
$\gamma_\mu \to \gamma_\mu \gamma_5$.
The remaining two currents correspond to the non-local symmetry transformations in Table~\ref{t:Sym}. 
Applying the Noether procedure and appealing to the equations of motion
yields
the isovector vector current 
\begin{eqnarray}
J^3_\mu(x)
&=&
\frac{1}{2}
\Big[
\delta_{\mathcal{B}} \, \ol \psi_x \gamma_\mu \delta_{\mathcal{B}} \, \psi_{x + \mu}
+
\delta_{\mathcal{B}} \, \ol \psi_{x+\mu} \gamma_\mu \delta_{\mathcal{B}} \, \psi_{x}
-
\delta_{\overline{\mathcal{B}}} \, \ol \psi_x \gamma_\mu \delta_{\overline{\mathcal{B}}} \, \psi_{x + \mu}
-
\delta_{\overline{\mathcal{B}}} \, \ol \psi_{x+\mu} \gamma_\mu \delta_{\overline{\mathcal{B}}} \, \psi_{x}
\notag \\
&& 
- i \lambda \delta_{\mu j} 
\Big(
\delta_{\mathcal{B}} \, \ol \psi_x \gamma_4 \delta_{\mathcal{B}} \, \psi_{x + j}
-
\delta_{\mathcal{B}} \, \ol \psi_{x + j} \gamma_4 \delta_{\mathcal{B}} \, \psi_x
-
\delta_{\overline{\mathcal{B}}} \, \ol \psi_x \gamma_4 \delta_{\overline{\mathcal{B}}} \, \psi_{x + j}
+
\delta_{\overline{\mathcal{B}}} \, \ol \psi_{x + j} \gamma_4 \delta_{\overline{\mathcal{B}}} \, \psi_x
\Big)
\Big],
\notag \\
\label{eq:IVV}
\end{eqnarray}
and the isosinglet axial current, 
$J_{\mu5}(x)$,
which is given by an expression identical to 
$J^3_\mu(x)$,
but with the replacement:
$\gamma_\mu \to \gamma_\mu \gamma_5$.

\section{Minimally doubled fermions with gauge interactions}         %
\label{s:VAVC}                                                                                  %

\subsection{Gauged action and its symmetries}  %

To include interactions with gauge fields, 
we gauge the isosinglet vector symmetry. 
On the lattice, 
the gauge fields are described by unitary link variables,
$U_{\mu,x}$. 
In the presence of such a gauge background, 
the Wilczek action becomes
\begin{eqnarray}
S [U]
&=&
\frac{1}{2}
\sum_{x, \mu}
\Big[
\ol \psi_x
U_{\mu,x}
\gamma_\mu 
\psi_{x + \mu} 
-
\ol \psi_{x + \mu}
U^\dagger_{\mu,x}
\gamma_\mu 
\psi_x 
\Big]
\notag \\
&& \phantom{space}
- 
\frac{i \lambda}{2} 
\sum_{x,j}
\Big[
\ol \psi_x
U_{j,x}
\gamma_4 
\psi_{x + j} 
+ 
\ol \psi_{x + j}
U^\dagger_{j,x}
\gamma_4 
\psi_x 
-
2
\ol \psi_x 
\gamma_4
\psi_x
\Big]
\label{eq:gauged}
.\end{eqnarray}
The gauge theory
retains only a subset of the symmetries possessed by the free theory. 
The non-local transformations listed in Table~\ref{t:Sym} no longer lead to symmetries because the link variables can carry momentum. 
Non-overlapping momentum regions are now connected by intermediate gluons, 
which can change quark flavor and chirality if the gluon energy is near 
$\pi$. 
Symmetry transformations in the gauge theory are best described in coordinate space. 
Aside from cubic invariance and gauge invariance, 
the symmetries of the gauge theory are collected in Table~\ref{t:SymU}.
While the continuous time reversal 
$\times$ 
flavor rotation symmetry is broken by gauge interactions, 
the particular flavor rotation by 
$\theta = \frac{\pi}{2}$
yields a discrete symmetry that we call flavored time reversal. 
%
\begin{table}[t]
\caption{%
Symmetries of the gauged, minimally doubled action. 
The coordinate-space transformations of the fermion and link fields are listed, 
along with the corresponding transformations of the free momentum-space fermion field. 
}
\label{t:SymU}
\begin{center}
\begin{tabular}{l||cc|c}
Gauge Theory Symmetry &
\multicolumn{2}{c|}{Coordinate Space} &
Free Theory Momentum Space 
\\
&
$\psi_x \longrightarrow$ &
$U_{\mu, x} \longrightarrow$
&
$\Psi (k) \longrightarrow$
\tabularnewline
\hline
\hline
Site Reflection ($\Xi$)
&
$ T \, \ol \psi {}^T_{\bm{1} - \bm{x}, x_4}$
&
$\left( U_{j, \bm{1} - \bm{x} - \hat{j}, x_4}, U^\dagger_{4, \bm{1} - \bm{x}, x_4}  \right)$ 
&
$ e^{ - i \sum_j k_j} (T \otimes 1) \ol \Psi \, {}^T(\bm{k}, -k_4)$ 
\\
Na\"ive Parity ($\mathcal{P}$)
&
$P  \psi_{- \bm{x}, x_4}$
&
$\left(U^\dagger_{j, - \bm{x} - \hat{j}, x_4}, \, U_{4, - \bm{x}, x_4} \right)$
&
$(P \otimes \tau^3) \Psi(-\bm{k}, k_4)$ 
\\
Na\"ive $\mathcal{C} \mathcal{T}$ &
$C T \, \ol \psi {}^T_{\bm{x}, -x_4}$
&
$\left( U^\dagger_{j,\bm{x}, - x_4}, U_{4, \bm{x}, -x_4 -1} \right)$
&
$(C T \otimes \tau^3) \ol \Psi {}^T (\bm{k}, - k_4)$ \\
Flavored Time Reversal ($\mathcal{F}$)
&
$e^{ i \pi x_4}  \psi_{\bm{x}, -x_4}$&
$\left(U_{j, \bm{x}, - x_4}, \, U^\dagger_{4, \bm{x}, -x_4-1} \right)$
&
$\quad ( T \otimes i \tau^2) \Psi( \bm{k}, - k_4) \quad$ 
\\
\hline
$U(1)$ Isosinglet Vector 
&
$ e^{ i \theta} \psi_x$
&
$U_{\mu, x}$
&
$e^{ i \theta ( 1 \otimes 1)} \Psi(k)$
\\
$U(1)$
Isovector Axial
&
$e^{i \theta \gamma_5} \psi_x$
&
$U_{\mu,x}$ 
&
$e^{ i \theta ( \gamma_5 \otimes \tau^3)} \Psi(k)$
\end{tabular}
\end{center}
\end{table}
Notice that the na\"ive parity transformation, 
$\mathcal{P}$,
given by applying the parity operation to the lattice field, 
$\psi_x$, 
yields a symmetry of the interacting theory. 
This na\"ive parity is not simply a parity transformation on each of the fermion flavors, 
but includes an additional isovector transformation by 
$\tau^3$ 
as well.
A similar situation is encountered for the na\"ive charge conjugation 
$\times$
time reversal transformation, 
$\mathcal{C} \mathcal{T}$. 
The transformation in Table~\ref{t:SymU} for the lattice field 
$\psi_x$, 
corresponds to a 
$CT$ 
transformation in addition to an isovector rotation by 
$\tau^3$.

To arrive at an interacting theory with the correct continuum limit, 
we must augment the action%
\footnote{
In practice, 
of course, 
one adds the lattice versions of the terms in Eq.~\eqref{eq:Sc}.
For example, 
the term
$i \ol \Psi ( \gamma_4 \otimes \tau^3) \Psi$
in the Symanzik action translates to the lattice operator 
$i \ol \psi_x \gamma_4 \psi_x$. 
The differences are of at least
$\mathcal{O}(a)$,
and have no consequence on our computation of the chiral anomaly in the continuum limit.
}
with the following relevant and marginal terms~%
\cite{Wilczek:1987kw,Bedaque:2008xs}
\begin{equation}
\cL_{c}
=
i c_0 \, \ol \Psi (\gamma_4  \otimes \tau^3 ) \Psi 
+
c_1 \, \ol \Psi D_4 ( \gamma_4 \otimes 1) \Psi
+
c_2 \, F_{4j} F_{4j}
\label{eq:Sc}
.\end{equation}
These Symanzik-level operators are allowed by the symmetries of the interacting theory.
We assume the coefficient 
$c_0$
has been fine-tuned to remove the power divergence in the continuum limit, 
and that
$c_1$ 
and 
$c_2$
have also been tuned to recover 
$SO(4)$
invariance.

As the local, continuous transformations remain symmetries of the gauge theory, 
see Table~\ref{t:SymU}, 
the corresponding symmetry currents are merely the gauged versions of Eq.~\eqref{eq:isosingletvector}. 
Applying the finite difference operator to the gauged symmetry currents, 
we find operator versions of the Ward-Takahashi identities. 
In an abritrary gauge background, 
the expectation value of these divergences vanish, 
$\sum_\mu  \langle  \nabla^*_\mu J_\mu (x) \rangle = \sum_\mu \langle \nabla^*_\mu J^3_{\mu 5} (x) \rangle = 0$.

\subsection{Non-conserved currents}    %

The Noether procedure shows us the isovector vector and isosinglet axial currents have non-vanishing divergences, 
namely
$\sum_\mu \,
\nabla^*_\mu
J^3_\mu(x) 
= 
D_V^3(x)$,
and
$\sum_\mu \,
\nabla^*_\mu
J_{\mu5}(x) 
= 
D_A(x)$,
with the currents
$J^3_\mu(x)$ 
and 
$J_{\mu 5}(x)$
given by the gauged versions of Eq.~\eqref{eq:IVV}, 
and their divergences, 
$D^3_V(x)$ 
and 
$D_A(x)$,  
having the form
\begin{eqnarray}
D_V^3 (x)
&=&
\left(
\frac{\delta S[ U]}{\delta \psi_x} 
\right)_{\overline{\mathcal{B}}}
\delta_{\mathcal{B}} \, \psi_x
-
\left(
\frac{\delta S[U]}{\delta \psi_x} \right)_\cB
\delta_{\overline{\mathcal{B}}} \, \psi_x
\notag \\
&& \phantom{space}
+ 
\delta_{\overline{\mathcal{B}}} \, \ol \psi_x  
\left(
\frac{\delta S [U] }{\delta \ol \psi_x} \right)_\cB
-
\delta_{\mathcal{B}} \, \ol \psi_x 
\left(
\frac{\delta S[U]}{\delta \ol \psi_x}
\right)_{\ol \cB}
\label{eq:DV}
, \\
D_A (x)
&=&
\left(
\frac{\delta S[U]}{\delta \psi_x} 
\right)_{\ol \cB}
\gamma_5
\, \delta_{\mathcal{B}} \, \psi_x
-
\left(
\frac{\delta S[U]}{\delta \psi_x} 
\right)_\cB
\gamma_5
\, \delta_{\overline{\mathcal{B}}} \, \psi_x
\notag \\
&& \phantom{space}
- 
\delta_{\overline{\mathcal{B}}} \, \ol \psi_x  
\gamma_5
\left( 
\frac{\delta S [U] }{\delta \ol \psi_x}
\right)_{\cB}
+
\delta_{\mathcal{B}} \, \ol \psi_x 
\, \gamma_5
\left(
\frac{\delta S[ U]}{\delta \ol \psi_x}
\right)_{\ol \cB}
\label{eq:DA}
.\end{eqnarray}
In the interacting theory, 
the temporal smearing is made gauge invariant by appending a Wilson line,
$\mathcal{W}_4 (\bm{x},x_4; \bm{x}, y_4)$, 
in the definition, Eq.~\eqref{eq:smear}. 
This Wilson line is the product of links in the time direction connecting the times 
$y_4$ 
and 
$x_4$. 
Derivatives of gauged action appearing above have the form:
$\left( \delta S[U] / \delta \ol \psi_x \right)_\cB = \sum_{x'} D(x,x') \delta_\cB \, \psi_{x'}$, 
for example,
where 
$D(x,x')$ 
denotes the coordinate-space Dirac operator of the interacting theory.   
Notice we have the relations, 
$\delta_\cB f_x + \delta_{\ol \cB} f_x = f_x$,
and
$\sum_x g_x ( \delta_\cB f_x ) = \sum_x  (\delta_\cB g_x)  f_x$, 
for any functions 
$f_x$ 
and 
$g_x$. 

%
\begin{table}[t]
\caption{%
Symanzik operator analysis of the anomalous divergences,  
$D_V^3$ 
and   
$D_A$.
Transformation properties of the divergences, 
as well as the relevant and marginal operators in the Symanzik effective theory are listed. 
For simplicity, 
we have suppressed all coordinate dependence by listing the transformations as either even or odd.
The coordinates of all operators transform in a covariant manner under the listed symmetries.
We have listed operators that are covariant under cubic rotations, 
and invariant under 
$U(1)_V$.
Operators not invariant under the isovector
$U(1)_A$
must be accompanied by an explicit power of the quark mass, 
$m_q$. 
The arrowed covariant derivative is defined by
$\overset{\leftrightarrow}{D} = \overset{\leftarrow}{D} {}^* - \overset{\rightarrow}{D}$, 
with 
$D_\mu = \partial_\mu + i A_\mu$. 
Notice we have utilized the equations of motion to eliminate $\Dslash \,$ terms.  
}
\label{t:Operators}
\begin{center}
\begin{tabular}{c||cccc}
 &
\multicolumn{4}{c}{Transformation} 
\\
Operator & 
$\quad \Xi \quad $ &
$\quad \mathcal{P} \quad$ &
$\quad \mathcal{C} \mathcal{T} \quad$ & 
$\quad \mathcal{F} \quad $ \\
\hline
\hline
$D_V^3$
&
$-$
&
$+$
&
$-$
&
$-$
\\
\hline
$\ol \Psi (\gamma_4 \otimes 1) \Psi$
&
$-$
&
$+$
&
$+$
&
$-$
\\
$\ol \Psi (\gamma_4 \otimes \tau^3) \Psi$
&
$-$
&
$+$
&
$+$
&
$+$
\\
$i \ol \Psi  \overset{\leftrightarrow}{D}_4  (\gamma_4 \otimes 1 ) \Psi$
&
$-$
&
$+$
&
$+$
&
$+$
\\
$i \ol \Psi  \overset{\leftrightarrow}{D}_4 (\gamma_4 \otimes \tau^3) \Psi$
&
$-$
&
$+$
&
$+$
&
$-$
\\
$D_4 [\, \ol \Psi (\gamma_4 \otimes 1 ) \Psi]$
&
$-$
&
$+$
&
$-$
&
$+$
\\
$D_4 [ \,\ol \Psi (\gamma_4 \otimes \tau^3) \Psi]$
&
$-$
&
$+$
&
$-$
&
$-$
\\
$i m_q \ol \Psi (1 \otimes 1) \Psi$
&
$-$
&
$+$
&
$+$
&
$+$
\\
$i m_q \ol \Psi (1 \otimes \tau^3) \Psi$
&
$-$
&
$+$
&
$+$
&
$-$
\\
$i F_{\mu \nu} F_{\mu \nu}$
&
$-$
&
$+$
&
$+$
&
$+$
\\
$i F_{4 \mu} F_{4 \mu}$
&
$-$
&
$+$
&
$+$
&
$+$
\\
\hline
$D_A$
&
$-$
&
$-$
&
$-$
&
$-$
\\
\hline
$ \ol \Psi (\gamma_4 \gamma_5 \otimes 1) \Psi$
&
$-$
&
$-$
&
$+$
&
$+$
\\
$ \ol \Psi (\gamma_4 \gamma_5 \otimes \tau^3) \Psi$
&
$-$
&
$-$
&
$+$
&
$-$
\\
$i \ol \Psi  \overset{\leftrightarrow}{D}_4  (\gamma_4 \gamma_5 \otimes 1 ) \Psi$
&
$-$
&
$-$
&
$+$
&
$-$
\\
$i \ol \Psi \overset{\leftrightarrow}{D}_4  (\gamma_4 \gamma_5 \otimes \tau^3) \Psi$
&
$-$
&
$-$
&
$+$
&
$+$
\\
$D_4 [\, \ol \Psi  (\gamma_4 \gamma_5 \otimes 1 ) \Psi]$
&
$-$
&
$-$
&
$-$
&
$-$
\\
$ D_4  [\, \ol \Psi (\gamma_4 \gamma_5 \otimes \tau^3) \Psi ]$
&
$-$
&
$-$
&
$-$
&
$+$
\\
$m_q \ol \Psi (\gamma_5 \otimes 1) \Psi$
&
$-$
&
$-$
&
$-$
&
$-$
\\
$m_q \ol \Psi (\gamma_5 \otimes \tau^3) \Psi$
&
$-$
&
$-$
&
$-$
&
$+$
\\
$\epsilon_{\mu \nu \a \b} F_{\mu \nu} F_{\a \b}$
&
$-$
&
$-$
&
$-$
&
$-$
\\
\end{tabular}
\end{center}
\end{table}
All operators in the divergences, Eqs.~\eqref{eq:DV} and \eqref{eq:DA},
connect different energy regions:
$\mathcal{B}$ 
to 
$\overline{\mathcal{B}}$,
and vice versa. 
Non-conservation of the isovector vector and isosinglet axial currents is hence due to flavor mixing. 
Without interactions, 
these divergences vanish due to flavor conservation in the non-interacting theory. 
Near the continuum limit, 
we can deduce the form of the anomalous divergences using the Symanzik effective action%
~\cite{Symanzik:1983dc}. 
In this limit, 
the divergences,
$D^3_V(x)$ 
and 
$D_A(x)$,
can be written in terms of continuum operators.
The tower of such operators can be ordered by their dimension, 
with the lowest dimension operators as the most relevant in the continuum limit. 
Our concern is only with those operators that survive when the lattice spacing, 
$a$, 
is taken to zero.

In Table~\ref{t:Operators}, 
we list the transformation properties of the divergences, 
$D_V^3$ 
and 
$D_A$, 
under the discrete symmetries of the lattice theory. 
We also list all possible relevant and marginal operators classified as to their transformation properties. 
From the table, 
we see that the divergence of the isovector vector current is required to have the form
\begin{equation} \label{eq:DVSymanzik}
D_V^3 
= 
c_V
D_4 [ \,\ol \Psi (\gamma_4 \otimes \tau^3) \Psi]
+ 
\mathcal{O}(a)
.\end{equation}
In other words, 
conservation of the isovector vector current in the continuum limit requires the fine tuning of a marginal operator. 
As the marginal operator breaks 
$SO(4)$ 
invariance, 
one might think that tuning the action to recover 
$SO(4)$
invariance in the continuum, 
via the terms in Eq.~\eqref{eq:Sc}, 
would be enough to render 
$c_V = 0$. 
This, 
however, 
is not the case. 
While the divergence of the isovector vector current depends on functional derivatives of the renormalized action, 
Eq.~\eqref{eq:DV},
there is additional 
$SO(4)$
breaking necessitated by the temporal non-locality of flavors. 
Even with a more local definition of the isovector vector current, 
ultimately 
$SO(4)$ 
must be broken, 
and fine tuning will be required so that 
$D_V^3(x) = 0$
in the continuum.

The divergence of the isosinglet axial current is required by discrete symmetries to have the form
\begin{equation}
D_A
= 
c_{A1}  \, 
m_q  \ol \Psi ( \gamma_5 \otimes 1) \Psi
+ 
c_{A2} \,
\epsilon_{\mu \nu \a \b} F_{\mu \nu} F_{\a \b}
+
c_{A3} \,
D_4 [ \, \ol \Psi ( \gamma_4 \gamma_5 \otimes 1 ) \Psi ]
.\end{equation}
The coefficient of the first term, 
is fixed by operator algebra,
$c_{A1} = 2$, 
while the coefficient of the second term, 
$c_{A2}$, 
should be fixed by the chiral anomaly. 
We pursue the computation of the anomaly in the next section.  
The last term is a surface term not excluded by the lattice symmetries, 
and modifies the divergence of the axial current in the chiral limit.
This 
$SO(4)$-breaking 
operator is similar to that encountered in the divergence of the isovector vector current, 
Eq.~\eqref{eq:DVSymanzik}.
Even with an action fine tuned to recover 
$SO(4)$
in the continuum, 
the coefficient 
$c_{A3}$
will not in general vanish.
This is due to 
$SO(4)$
breaking introduced by the temporal non-locality required to form the isosinglet flavor combination in the axial current.  
Alternate definitions of the current must share this malady, 
and one must fine tune the current to arrive at 
$c_{A3} = 0$. 
We will assume this tuning has been performed, 
so that the operator
$\epsilon_{\mu \nu \a \b} F_{\mu \nu} F_{\a \b}$ 
(or $\epsilon_{\mu \nu} F_{\mu \nu}$ in two dimensions)
is the only contribution to 
$D_A$
in the chiral and continuum limits.

\section{Chiral Anomaly}                                                                %
\label{s:CA}                                                                                    %

For perturbatively weak gauge backgrounds, 
we can compute the divergence of the isosinglet axial current
using lattice perturbation theory. 
It is useful to employ a compact notation;
and, 
to this end,
we write the minimally doubled action in the form,
$S = \sum_{x,y} \ol \psi_x \langle x | D | y \rangle \psi_y$,
so that, 
as an abstract operator, 
the propagator is merely 
$D^{-1}$.  
From Eq.~\eqref{eq:DA},
the expectation value of the axial current's divergence in a gauge background is
\begin{eqnarray}
\langle D_A(x) \rangle
=
\text{Tr} \,
\gamma_5
\langle x | 
\Big[
D^{-1}_{\ol \cB  \cB},
D
\Big] 
| x \rangle
-
( \cB \leftrightarrow \ol \cB )
,\end{eqnarray}
where we have used 
$\{ \gamma_5, D \} = 0$, 
and the trace is over spin. 
The flavor-changing propagators are given by Wick contractions of time-smeared fields
\begin{equation}
D_{\ol \cB \cB}^{-1} (x,x')
=
\langle \delta_{\ol \cB} \, \psi_x \, \delta_{\cB} \, \ol \psi_{x'} \rangle
,\end{equation}
with a similar expression for 
$D_{\cB \ol \cB}^{-1}(x,x')$. 
Expanding in powers of the external gauge field, 
we write
$D = D^{(0)} + D^{(1)} + \ldots$, 
where the ellipsis represents terms with at least two powers of the gauge field.

\begin{figure}
\epsfig{file=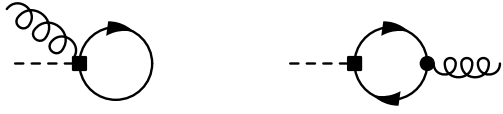,width=0.4\textwidth}
\caption{%
Leading contributions to the two-dimensional Abelian anomaly in lattice perturbation theory.
The squares denote insertion of the isosinglet axial current's divergence, 
while the circle denotes the gauge vertex. 
Fermions are directed lines, 
gauge fields are curly lines, 
and the flow of momentum from the axial current is shown with dashed lines.  
}
\label{f:anomaly}
\end{figure}

For simplicity, 
we restrict our attention to the Abelian anomaly in two dimensions. 
A further simplification is afforded by adopting Coulomb gauge, 
for which the time-links are unity. 
The leading contributions to the divergence appear at linear order in the gauge field. 
Thus we have
\begin{eqnarray}
\langle D_A(x) \rangle
&=&
\text{Tr}
\, 
\gamma_5
\langle x |
\Big[
D^{(0)^{-1}},
D_{\cB \ol \cB}^{(1)}
\Big]
x \rangle
+
\text{Tr} \,
\gamma_5
\langle x |
\Big[
D^{(0)^{-1}} D_{\cB \ol \cB}^{(1)} \, D^{(0)^{-1}},
D^{(0)}
\Big]
x \rangle
-
( \cB \leftrightarrow \ol \cB )
.\notag \\
\end{eqnarray}
The first trace corresponds the first diagram in Fig.~\ref{f:anomaly}, 
while the second trace corresponds to the second diagram. 
Both diagrams, however, are identical, and are easiest to evaluate in momentum space.

%
%
%
%

In momentum space, 
the gauge vertex, 
$\cM(k+q,k)$,
in Coulomb gauge is defined by
\begin{equation}
\sum_y e^{i q' \cdot y} \frac{\delta}{\delta A_1(y)} \langle k+q | D^{(1)} | k \rangle
\equiv
(2 \pi)^2 \delta( q' - q) \cM (k+q, q)
.\end{equation}
Using the Wilczek action, 
we have
\begin{equation}
\cM (k+q, k) 
=
\frac{i e}{2} \gamma_1
[ e^{ i k_1} + e^{ - i (k_1 + q_1) }]
+ 
\frac{\lambda e}{2} \gamma_2
[ e^{ i k_1} - e^{ - i (k_1 + q_1) }]
.\end{equation}
The free Wilczek propagator in momentum space has the form
\begin{equation}
D^{(0)^{-1}} (k) 
= 
- D^{(0)}(k) \, \cD^{-1}(k)
,\end{equation}
with 
$D^{(0)} (k) = i \gamma_1 \sin k_1 + i \gamma_2 [ \sin k_2 - \lambda ( \cos k_1 - 1)]$, 
which is the 
two-dimensional version of Eq.~\eqref{eq:free},
and
$\cD(k) = \sin^2 k_1 + \left[ \sin k_2 - \lambda ( \cos k_1 - 1) \right]^2$.
Injecting momentum 
$q$ 
into the divergence, 
we have the momentum-space result
\begin{eqnarray}
\langle D_A(q) \rangle
&=&
2 
\underset{k\in \cB, \, k + q \in \ol \cB}{\int}
\frac{d^2 k}{(2 \pi)^2}
\text{Tr} \,
\gamma_5 \,
\Big[ D^{(0)^{-1}} (k) + D^{(0)^{-1}} (k+q) \Big]
\cM (k+q,k)
- ( \cB \leftrightarrow \ol \cB)
\label{eq:midstep}
.\end{eqnarray}

Contributions to Eq.~\eqref{eq:midstep} around 
$q_\mu = (0,0)$
appear only at second order in 
$q$. 
The minimally doubled action,
however, 
possesses two low-energy regions, 
and there are linear contributions around the momentum
$q_\mu = (0, \pi)$. 
Writing 
$q_\mu \to \pi \delta_{\mu 2} + q'_\mu$, 
we then expand  
$q'_\mu$ 
about zero retaining all linear terms.  
First, 
we notice that restrictions on the integration region allow us to split the integration into a finite region and an infinitesimally small region, 
\begin{eqnarray}
\underset{k\in \cB, \, k + q' \in \cB}{\int}
= 
\underset{k\in \cB}{\int}
- 
\underset{k\in \cB, \, k + q' \in \ol \cB}{\int}
.\end{eqnarray}
The integral over the infinitesimal region can be shown to vanish quadratically in 
$q'_\mu$. 
Second,
we notice that in the remaining finite interval, 
the integrand is odd in 
$k_2$ 
if 
$q'_2 = 0$.
Thus the only linear contribution is proportional to 
$q'_2$;
and, 
after some algebra has the form
\begin{eqnarray}
\langle D_A(q) \rangle
&=&
- 4 (q_2 - \pi) A_1(q)
\underset{k\in \cB}{\int}
\frac{d^2 k}{(2 \pi)^2}
\text{Tr} \,
\gamma_5 \,
\frac{ \partial D^{(0)^{-1}} (k)}{\partial k_2}    
\cM (k, k)
\label{eq:midstep2}
.\end{eqnarray}
A valuable simplification is shifting the energy integration by 
$\pi$
in the 
$(\cB \leftrightarrow \ol \cB)$ 
term. 
At this stage, 
we utilize 
$\text{Tr} \, \gamma_5 \gamma_\mu \gamma_\nu = 2 i \epsilon_{\mu \nu}$ 
to perform the spin trace. 
The divergence of the axial current then takes the form
\begin{eqnarray}
\langle D_A(q) \rangle
&=&
- 8i  e (q_2 - \pi) A_1(q)
\underset{k\in \cB}{\int}
\frac{d^2 k}{(2 \pi)^2}
\frac{\partial}{\partial k_2}
\Bigg(
\frac{
\lambda \sin^2 k_1 - \cos k_1 [ \sin k_2 - \lambda ( \cos k_1 - 1) ]
}{\cD(k)}
\Bigg)
\notag \\
&=&
- 8i e (q_2 - \pi) A_1(q)
\underset{k\in \cB}{\int}
\frac{d^2 k}{(2 \pi)^2}
\frac{\partial}{\partial k_1}
\left( 
\frac{\sin k_1 \cos k_2}{\cD(k)}
\right)
\label{eq:anomaly}
.\end{eqnarray}
As the surface terms vanish, 
the only contribution to the divergence arises from the region where
$|k_1|$, $|k_2| < \varepsilon$, 
with 
$\varepsilon \to 0$. 
The integrand has an integrable singularity in this region. 
Carrying out the integration for finite 
$\varepsilon$,
and then taking the limit 
$\varepsilon \to 0$
produces the correct chiral anomaly
\begin{equation}
\langle D_A(q) \rangle
=
- \frac{2 i e}{\pi} (q_2 - \pi) A_1(q)
\longrightarrow
- \frac{e N_f}{2 \pi} \epsilon_{\mu \nu} F_{\mu \nu}, 
\quad 
\text{with }
N_f = 2 
.\end{equation}

\section{Minimally Doubled Weyl Fermions}                                  %
\label{s:MDWF}                                                                              %

\subsection{Free Action}%

Using the chiral projector, 
$P_R = \frac{1}{2} ( 1 + \gamma_5 )$, 
on a minimally doubled action,
we can reduce the number of spin components by half. 
The resulting action can be written in terms of a two-component Weyl spinor
$\chi_x$, 
namely
\begin{eqnarray} 
S 
&=& 
\frac{1}{2}
\sum_{x, \mu}
\Big[
\chi^\dagger_x
i \sigma_\mu 
\chi_{x + \mu} 
-
\chi^\dagger_{x + \mu}
i \sigma_\mu 
\chi_x 
\Big]
- 
\frac{i \lambda}{2} 
\sum_{x, j}
\Big[
\chi^\dagger_x 
\chi_{x + j} 
+ 
\chi^\dagger_{x + j}
\chi_x 
-
2
\chi^\dagger_x 
\chi_x
\Big]
\label{eq:Waction}
,\end{eqnarray}
with 
$\sigma_\mu = ( \bm{\sigma}, - i )$. 
There are again two low-energy modes, 
which we now label by
$k_\mu^{(R)} = (0,0,0,0)$
and
$k_\mu^{(L)} = (0,0,0,\pi)$. 
Fields with chirality can be created by partitioning the Brilluoin zone.
We take
$\chi(k) \Big|_{k_\mu \in \mathcal{B}}
=
\chi^{(R)}(k)$,
for the right-handed field, 
and 
$\chi(k) \Big|_{k_\mu \notin \mathcal{B}}
= 
\chi^{(L)} (T_{\pi 4} k)$,
for the left-handed field.
Forming a Dirac spinor from the left- and right-handed fields, 
$\psi(k) = \begin{pmatrix} \chi^{(R)} (k) \\ \chi^{(L)} (k) \end{pmatrix}$, 
and its conjugate spinor,
$\ol \psi(k) = \Big( - \chi^{\dagger (L)} (k), \chi^{\dagger (R)} (k) \Big)$, 
we can write the action in the form
\begin{eqnarray}
S
&=&
\int_{\mathcal{B}} \frac{d^4 k}{(2 \pi)^4}
\ol \psi(k)
\Big[
\sum_\mu i \gamma_\mu \sin k_\mu
- i \lambda \gamma_4 \gamma_5
\sum_j (\cos k_j -1)
\Big]
\psi(k)
,\end{eqnarray}
with Euclidean chiral basis Dirac matrices 
$\gamma_\mu = \begin{pmatrix} 0 & - i \ol \sigma_\mu \\ i \sigma_\mu & 0 \end{pmatrix}$, 
$\ol \sigma_\mu = ( \bm{\sigma}, i )$,
and 
$\gamma_5 = \diag ( 1 , -1 )$.

The free momentum-space action has:
cubic symmetry;
invariance under the product of charge conjugation and parity, $CP$;
time reversal invariance, $T$;
$U(1)$ vector symmetry, $e^{ i \theta}$;
$U(1)$ axial symmetry, $e^{ i \theta \gamma_5}$. 
In coordinate space, 
the continuous vector symmetry is local, 
while the axial symmetry is non-local in time
\begin{equation}
\chi_x
\to 
\begin{cases}
e^{ i \theta} \chi_x, 
\quad &
U(1)_V \\
e^{ i \theta}
\delta_{\mathcal{B}} \, \chi_x
+
e^{ - i \theta}
\delta_{\overline{\mathcal{B}}} \, \chi_x, 
\quad  &
U(1)_A
\end{cases}
\label{eq:symms}
.\end{equation}
Thus the lattice current corresponding to the 
$U(1)_V$
symmetry, 
\begin{eqnarray}
J_\mu (x)
&=& 
\frac{1}{2}
\Big[
\chi^\dagger_x
i \sigma_\mu 
\chi_{x + \mu} 
+
\chi^\dagger_{x + \mu}
i \sigma_\mu 
\chi_x 
- 
i \lambda \delta_{\mu j} 
\Big(
\chi^\dagger_x 
\chi_{x + j} 
- 
\chi^\dagger_{x + j}
\chi_x 
\Big)
\Big]
\label{eq:WV}
,\end{eqnarray}
is conserved, 
while that corresponding to the non-local 
$U(1)_A$
symmetry, 
\begin{eqnarray}
J_{\mu5} (x)
&=&
\frac{1}{2}
\Big[
\delta_{\mathcal{B}} \, \chi^\dagger_x i \sigma_\mu \delta_{\mathcal{B}} \, \chi_{x + \mu}
+
\delta_{\mathcal{B}} \,  \chi^\dagger_{x+\mu} i \sigma_\mu \delta_{\mathcal{B}} \, \chi_{x}
-
\delta_{\overline{\mathcal{B}}} \, \chi^\dagger_x i \sigma_\mu \delta_{\overline{\mathcal{B}}} \, \chi_{x + \mu}
-
\delta_{\overline{\mathcal{B}}} \, \chi^\dagger_{x+\mu} i \sigma_\mu \delta_{\overline{\mathcal{B}}} \, \chi_{x}
\notag \\
&& 
- i \lambda \delta_{\mu j} 
\Big(
\delta_{\mathcal{B}} \, \chi^\dagger_x \delta_{\mathcal{B}} \, \chi_{x + j}
-
\delta_{\mathcal{B}} \, \chi^\dagger_{x + j}  \delta_{\mathcal{B}} \, \chi_x
-
\delta_{\overline{\mathcal{B}}} \, \chi^\dagger_x  \delta_{\overline{\mathcal{B}}} \, \chi_{x + j}
+
\delta_{\overline{\mathcal{B}}} \, \chi^\dagger_{x + j} \delta_{\overline{\mathcal{B}}} \, \chi_x
\Big)
\Big],
\label{eq:WA}
\end{eqnarray}
is only conserved in the free theory.

\subsection{Gauge Theory and Chiral Anomaly} %

The inclusion of gauge fields is straightforward; 
one gauges the 
$U(1)_V$ 
symmetry of the free theory.
As in the case of the Wilczek action, 
additional relevant and marginal terms must be added with fine-tuned coefficients in order to arrive at the correct continuum limit. 
As the terms needed are similar to those given above in Eq.~\eqref{eq:Sc}, 
we do not spell them out explicitly.

The vector symmetry transformation in Eq.~\eqref{eq:symms} acts locally on the fermion field, 
it remains a symmetry of the interacting theory. 
The corresponding vector current is the gauged version of Eq.~\eqref{eq:WV}, 
and has vanishing divergence in an arbitrary gauge field background.  
The same fate does not befall the axial symmetry in the presence of gauge interactions. 
The gauged axial current in Eq.~\eqref{eq:WA} has non-vanishing divergence
\begin{eqnarray}
\sum_\mu \nabla^*_\mu  J_{\mu 5}(x)
&=&
\left(
\frac{\delta S[ U]}{\delta \chi_x} 
\right)_{\overline{\mathcal{B}}}
\delta_{\mathcal{B}} \, \chi_x
-
\left(
\frac{\delta S[U]}{\delta \chi_x} \right)_\cB
\delta_{\overline{\mathcal{B}}} \, \chi_x
+ 
\delta_{\overline{\mathcal{B}}} \, \chi^\dagger_x  
\left(
\frac{\delta S [U] }{\delta \chi^\dagger_x} \right)_\cB
-
\delta_{\mathcal{B}} \, \chi^\dagger_x 
\left(
\frac{\delta S[U]}{\delta \chi^\dagger_x}
\right)_{\ol \cB}
.\notag \\
\label{eq:DWA}
\end{eqnarray}
Left-handed and right-handed fields mix due to interactions with gauge fields, 
and this mixing is the source of the chiral anomaly.

Due to the breaking of 
$SO(4)$
invariance that is inherent to defining the axial current, 
Eq.~\eqref{eq:WA},
fine tuning is required to recover the chiral anomaly in the continuum limit.
At the Symanzik level, 
the divergence of the axial current receives a contribution from an 
$SO(4)$ breaking operator, 
namely
$D_4 [ \, \ol \psi \gamma_4 \gamma_5 \psi]$. 
We assume that this contribution has been tuned to vanish by improving the current.
This leaves only a potential contribution to the divergence from the pure gauge operator
$\epsilon_{\mu \nu \a \b} F_{\mu \nu} F_{\a \b}$
(or, in two dimensions, 
$\epsilon_{\mu \nu} F_{\mu \nu}$).  
Using the divergence of the axial current, 
Eq.~\eqref{eq:DWA},
we can compute its expectation value in a gauge field background to determine the coefficient of the pure gauge operator. 
For simplicity, 
we work in 
two dimensions, 
for which the Weyl spinors are one-component objects, 
and
$\sigma_\mu = ( 1, -i)$.
In a weak gauge field, 
we arrive at the diagrams shown in Figure~\ref{f:anomaly}.
Their evaluation parallels that above in Sec.~\ref{s:CA}, 
and one encounters integrals identical to those arising in the computation of the axial anomaly, 
Eq.~\eqref{eq:anomaly}. 
Notably absent is the trace over spin indices, 
which reduces the overall value by a factor of two. 
Consequently we find, 
\begin{equation}
\sum_\mu 
\langle 
\nabla_\mu^* 
J_{\mu 5} 
\rangle
=
- \frac{e N_f}{2 \pi} \epsilon_{\mu \nu} F_{\mu \nu}, 
\quad
\text{with }
N_f = 1
.\end{equation}
Thus the continuum limit of the minimally doubled Weyl action,
Eq.~\eqref{eq:Waction},
describes a single Dirac fermion, 
and reproduces the chiral anomaly. 
Fine tuning of both the action and the axial current, 
however, 
is required to obtain this result.

\section{Summary}                                                                        %
\label{s:S}                                                                                      %

Through the investigation of minimally doubled fermions, 
we have explained the appearance of the axial anomaly
in chirally symmetric lattice actions. 
Although such free actions have a flavor-singlet axial symmetry, 
the lattice field transforms non-locally under this symmetry.
When gauge interactions are included, 
the symmetry is broken by flavor changing mediated by gluons carrying the requisite momentum. 
In the continuum limit, 
we demonstrated that these interactions produce the correct chiral anomaly in two dimensions.
Similar flavor-changing interactions lead to non-vanishing divergences of non-singlet vector currents.
To arrive at a conserved non-singlet vector current, 
fine tuning is required to restore 
$SO(4)$ 
invariance. 
The singlet axial current must also be tuned to recover
$SO(4)$. 
Finally we considered the chiral projection of minimally doubled fermion actions. 
Without gauge fields, 
these projected actions describe a single, 
chirally symmetric Dirac fermion. 
Chirality-changing interactions are present in the gauged theory;
they supply precisely the right contribution to the chiral anomaly.

\begin{acknowledgments}
We thank M.~I.~Buchoff for inciting this investigation. 
This work is supported in part by the 
U.S.~Dept.~of Energy,
under
Grant No.~DE-FG02-93ER-40762.
The Institute for Nuclear Theory at the University of Washington is acknowledged for its hospitality 
and partial support during the course of this work.
\end{acknowledgments}

\appendix%

\section*{Wilson Lattice Action}

It is straightforward to compute the chiral anomaly for the Wilson action in momentum space. 
We present a simple derivation. 
The Wilson action takes the form
\begin{eqnarray}
S_W [U]
&=&
\frac{1}{2}
\sum_{x, \mu}
\Big[
\ol \psi_x
U_{\mu,x}
(\gamma_\mu +  r)
\psi_{x + \mu} 
-
\ol \psi_{x + \mu}
U^\dagger_{\mu,x}
(\gamma_\mu - r)
\psi_x 
-
2 (r - m_0) \, \ol \psi_x \psi_x
\Big]
\label{eq:Wilson}
,\end{eqnarray}
where 
$r$ 
and
$m_0$ 
are real-valued parameters, 
the latter is the bare value of the fermion mass. 
Because of chiral symmetry breaking, 
the bare mass receives an additive renormalization proportional to
$\alpha_s / a$, 
where 
$a$ 
is the lattice spacing. 
Below, 
we implicitly tune the bare parameter
$m_0$
to exactly cancel the power divergent mass renormalization. 
This fine tuning is required to arrive at the correct anomaly in the continuum limit.

Even with vanishing renormalized quark mass, 
the axial transformation, 
$\psi_x \to e^{ i \theta \gamma_5} \psi_x$, 
is not a symmetry of the Wilson action. 
As a consequence, 
the axial-vector current, 
$J_{\mu 5}(x)$, 
given by
\begin{eqnarray}
J_{\mu 5}(x)
= 
\frac{1}{2} 
\Big[
\ol \psi_x U_{\mu,x} \gamma_\mu \gamma_5 \psi_{x +\mu}
+ 
\ol \psi_{x +\mu} U^\dagger_{x,\mu} \gamma_\mu \gamma_5 \psi_{x}
\Big]
,\end{eqnarray}
is not conserved. 
Writing the action as 
$S_W = \sum_{x,y} \ol \psi_x  \langle x | D_W | y \rangle \psi_y$, 
with the Wilson-Dirac operator defined by
$D_W = \Dslash + R$, 
casts the divergence in the form
\begin{equation}
\langle \nabla^*_\mu J_{\mu 5} (x) \rangle
=
- 
\text{Tr}
\,
\gamma_5
\langle x |
\Big\{
D_W^{-1}, 
R
\Big\}
| x \rangle
=
\text{Tr}
\,
\gamma_5
\langle x |
\Big\{
D^{-1}_W, 
\Dslash 
\Big\}
| x \rangle
.\end{equation}
Expanding in powers of the gauge coupling, 
we write
$D_\mu = D_\mu^{(0)} + D_{\mu}^{(1)} + \ldots$, 
and
$R = R^{(0)} + R^{(1)} + \ldots$. 
In two dimensions, 
the leading contribution to the divergence occurs at first order in the gauge field
\begin{equation}
\langle \nabla^*_\mu J_{\mu 5} (x) \rangle
=
\text{Tr}
\,
\gamma_5
\Big\langle x \Big|
\Big\{
D^{(0)^{-1}}_W, 
\Dslash \, {}^{(1)}
\Big\}
-
\Big\{
D_W^{(0)^{-1}}
( \Dslash \, {}^{(1)} + R^{(1)} )
D_W^{(0)^{-1}}, 
\Dslash \, {}^{(0)}
\Big\}
\Big| x \Big\rangle
+
\ldots
.\end{equation}
The first anti-commutator is graphically depicted as the first diagram in Fig.~\ref{f:anomaly}, 
while the second anti-commutator is depicted by the second diagram. 
In momentum space, 
the free Wilson propagator is
\begin{equation}
D_W^{(0)^{-1}} (k) 
= 
\sum_\mu [- i \gamma_\mu \sin k_\mu + r ( \cos k_\mu - 1) ]
\cD^{-1}(k)
,\end{equation}
with 
$\cD(k) = \sum_\nu \sin^2 k_\nu + r^2 \left[ \sum_\nu (\cos k_\nu -1) \right]^2$,
and the gauge interaction vertices are
\begin{eqnarray}
\delta \Dslash \, {}^{(1)}  / \delta A_\mu 
&\longrightarrow&
\cM_\mu (k+q, k) 
=
\frac{i e}{2} \gamma_\mu [ e^{ i k_\mu} + e^{i ( k_\mu + q_\mu)} ]
\\
\delta R^{(1)} / \delta A_\mu
&\longrightarrow& 
\cR_\mu (k+q,k)
=
\frac{i r e}{2} [ e^{ i k_\mu} - e^{i ( k_\mu + q_\mu)} ]
.\end{eqnarray}

Evaluating the traces, 
we find 
\begin{eqnarray}
\langle \nabla^*_\mu J_{\mu 5} (q) \rangle
&=&
2 i e \epsilon_{\mu \nu} q_\mu A_\nu
\int_{-\pi}^\pi \frac{d^2k}{( 2 \pi)^2}
\left[
\frac{\partial}{\partial k_\nu}
\left( 
\frac{\cos k_\mu \sin k_\nu}{\cD(k)}
\right)
+
\frac{\partial}{\partial k_\mu}
\left( 
\frac{\cos k_\nu \sin k_\mu}{\cD(k)}
\right)
\right]
,\end{eqnarray}
where we have kept the two terms ordered as to their corresponding diagrams. 
Non-vanishing contributions to the integral arise from the region,
$|k_\mu|$,  $|k_\nu| < \varepsilon $, 
where the integrand has an integrable singularity. 
Taking 
$\varepsilon \to 0$, 
we obtain
\begin{equation}
\langle \nabla^*_\mu J_{\mu 5} (q) \rangle
=
-
\frac{e N_f}{2 \pi}
\epsilon_{\mu \nu} F_{\mu \nu}(q), 
\quad 
\text{with }
N_f = 1
.\end{equation}

\bibliography{bibfile}%

\end{document}